\documentclass[11pt]{article}

\usepackage{amsfonts,amsmath,amssymb,enumerate,epsfig}
\usepackage{theorem}
\usepackage{appendix}
\usepackage{color}

\numberwithin{equation}{section}
\newtheorem{definition}{Definition}[section]

\newtheorem{proposition}[definition]{Proposition}
\newtheorem{theorem}[definition]{Theorem}

\newtheorem{lemma}[definition]{Lemma}

\newcommand{\prf}{\underline{Proof:}\ }
\newcommand{\finprf}{\null \hfill {\rule{5pt}{5pt}}\\ \null}
\newcommand{\ie}{{\it i.e.}\ }

\newcommand{\be}{\begin{equation}}
\newcommand{\ee}{\end{equation}}
\newcommand{\beu}{\begin{equation*}}
\newcommand{\eeu}{\end{equation*}}
\newcommand{\bea}{\begin{eqnarray}}
\newcommand{\eea}{\end{eqnarray}}
\newcommand{\beaa}{\begin{eqnarray*}}
\newcommand{\eeaa}{\end{eqnarray*}}
\newcommand{\bmx}{\begin{pmatrix}}
\newcommand{\emx}{\end{pmatrix}}

\def\cM{{\cal M}}

\newcommand{\CC}{{\mathbb C}}

\newcommand{\RR}{\mbox{${\mathbb R}$}}

\newcommand{\1}{\mbox{\hspace{.0em}1\hspace{-.24em}I}}

\advance\textheight by \topskip
\begin{document}

\renewcommand{\thefootnote}{\arabic{footnote}}
\setcounter{footnote}{0}
\newpage
\setcounter{page}{0}

\begin{center}

{\Huge \textbf{On the Inverse Scattering Method for Integrable PDEs on a Star Graph}}

\vspace{2cm}

{\Large \textbf{Vincent Caudrelier} }

\vspace{0.5cm}

\emph{\Large Department of Mathematics,
 City University London,\\
  Northampton Square, London EC1V 0HB, UK}

\vfill

\begin{abstract}
We present a framework to solve the open problem of formulating the inverse scattering method (ISM) for an integrable PDE on a star-graph. The idea is 
to map the problem on the graph to a matrix initial-boundary value (IBV) problem and then to extend the unified method of Fokas to such a matrix IBV 
problem. The nonlinear Schr\"odinger equation is chosen to illustrate the method. The framework unifies all previously known examples which are 
recovered as particular cases. The case of general Robin conditions at the vertex is discussed: 
the notion of linearizable initial-boundary conditions is introduced. For such conditions, the method is shown to be as efficient as the ISM on the full-line.

\end{abstract}
\end{center}
\vfill


\newpage
\pagestyle{plain}

\section{Introduction}

For decades now, integrable partial differential equations (PDEs), and more generally integrable systems, have fuelled research and important discoveries 
in Mathematics and Physics, and still do. Comparatively more recently, graphs and dynamical systems on (quantum) graphs have emerged as a successful 
framework to model a large variety of (complex) systems. It is therefore not surprising to see a fast growing interest in developing a theory of integrable 
systems on graphs, which would combine the power of integrable systems with the flexibility of graphs to model more realistic situations. 
 The review \cite{Noja}, for instance, gives a flavour and references for this fast growing area in the context of nonlinear Schr\"odinger (NLS) equations (not restricted to integrable cases).

Originally, integrable PDEs were treated as initial value problems for functions of one space variable $x\in\RR$ and one time variable $t\ge 0$. 
The invention 
of the inverse scattering method (ISM) \cite{GGKM} and its refinements \cite{ZS,AKNS} through the systematic use of a Lax pair \cite{Lax} represents a 
cornerstone of modern integrable PDEs. The first departure from this setup to solve an initial-boundary value (IBV) problem for an 
integrable PDE on the half-line \cite{AS,BT} or a finite interval \cite{Sk} can 
be viewed in retrospect as the beginning of the study of integrable PDEs on metric graphs. Indeed, a half-line is nothing but a half-infinite edge attached to 
a vertex and a finite interval is a finite edge connecting two vertices. The next big step in this natural evolution was the study of integrable 
PDEs on the line with a defect/impurity at a fixed site (or possibly several such defects). The vast literature on this problem\footnote{Given the 
large literature, we have tried to give a representative selection of papers related to classical systems with defects. Most of the authors 
cited here have contributed many more papers on the subject.}
\cite{BCZ,CZ,CMR,GYZ,C,AD,HK,DK,AAGZ,Ag}
shows both its interest and its difficulty. To date however, despite some impressive results on the behaviour of certain solutions \cite{HMZ,DP}, 
the general problem of formulating an ISM for a problem with defects is still open. 

It is the purpose of this paper to bring an answer to the more general question of formulating the ISM on a star-graph, 
\ie a single vertex connected to a finite number $N$ of half-infinite edges. The case $N=2$ will then take care of the 
situation of a defect/impurity on the line. 
To be more concrete, we choose to present the framework on the example of the cubic Nonlinear Schr\"odinger (NLS) equation 
\be
\label{eq:NLS}
i\partial_t q+\partial_x^2 q-2g|q|^2q=0\,,~~g\in\RR\,.
\ee
This is motivated by the fact that it is one of the most famous and 
studied example of integrable PDEs. It is also the model that has been most studied on various simple graphs, hence allowing us to show how our method 
encompasses all known results. But it will be clear to the reader that our framework applies equally well to any integrable PDE that admits a Lax pair
formulation.
\paragraph{Summary of the results:} The main results of this paper are Theorems \ref{theorem_mapping} and \ref{reconstruction_theorem}. 
Theorem \ref{theorem_mapping}, and its important consequence for the 
spectral analysis of the Lax pair formulation of an integrable PDE on a star-graph, basically ensures that all the tools developed within the 
unified transform method (see e.g. \cite{Fokas}) can be directly lifted to diagonal-matrix valued spectral data in order to analyse any integrable PDE on a star-graph 
that is analysable by the Fokas method on the half-line. Theorem \ref{reconstruction_theorem} shows, on the example of NLS, how the general procedure works
with the same level of generality as the analog procedure for models on the half-line. The key ingredient is the Riemann-Hilbert problem discussed 
in Section \ref{inverse_part}, which also serves to illustrate the distinctive features of the present matrix cases compared to the scalar case
$[$see conditions (C1)-(C3) in Section \ref{spectral_analysis}$]$. Finally, the role of the global relation for a star-graph problem is discussed and its 
solution via the Dirichlet-to-Neumann map is presented in Proposition \ref{solution_H1}.

In the next section, we introduce the model to solve: NLS on a star-graph. In section \ref{mapping}, we show how the problem can be mapped to 
an IBV problem of a certain matrix form. Section \ref{ISM} then goes on to exploit this mapping in combination with the unified method of Fokas 
\cite{Fokas} to provide an ISM for NLS on a star-graph. In Section \ref{compare}, we present a detailed example of 
linearizable boundary conditions for $N=2$ illustrating how the full-line problem is recovered from our setup. 
We also show how previous studies fit within our framework. Finally, 
in Section \ref{general_case}, we illustrate our approach in the general case of a star-graph with the vertex boundary conditions classified in 
\cite{KS}. Based on this example, we introduce the notion of linearizable initial-boundary conditions whereby one can reduce the problem on the star-graph 
to a scalar linearizable IBV problem. Conclusions are gathered in the last section, where future directions are also pointed out.

\section{The Model to Solve}

When formulated on the real line, \eqref{eq:NLS} is solved via the ISM for initial data $q(x,0)=q_0(x)$ typically of sufficiently fast decay (see e.g. 
\cite{FT} and references therein). Of course, other classes of initial data have been considered over the years (see e.g. \cite{Dem} and references therein). 
In this paper, we want to focus 
on the presentation of the new framework and not on the technicalities related to local or global well-posedness and the choice of appropriate functional spaces.
These important questions are beyond the scope of the present paper. We assume for now that all our data is of appropriate
smoothness and decay for our purposes. 

We consider the NLS equation on a star-graph with $N$ half-infinite edges. We introduce
$N$ copies of \eqref{eq:NLS} for functions $q^\alpha$, $\alpha=1,\dots,N$. Each $q^\alpha$ lives on edge $\alpha$, is a function of $x\ge 0$ and $t\ge 0$ and 
is connected to the other edges via some boundary conditions at $x=0$.
The problem therefore reads, for $\alpha=1,\dots,N$,
\bea
\label{NLS_edge}
&& i\partial_t q^\alpha+\partial_x^2 q^\alpha-2g|q^\alpha|^2q^\alpha=0~~,~~x,t>0\,,\\
&& \label{IC_edge} q^\alpha(x,0)=q_0^\alpha(x)\,,\\
&& \label{BC_edge} q^\alpha(0,t)=g_0^\alpha(t)\,,~~\partial_x q^\alpha(0,t)=g_1^\alpha(t)\,,
\eea
where $q_0^\alpha$, $g_0^\alpha$
and $g_1^\alpha$ are the initial-boundary data. For each $\alpha=1,\dots,N$, \eqref{NLS_edge} is the compatibility condition $\partial_{xt}\mu=\partial_{tx}\mu$ of \cite{ZS}
\be
\label{pb_aux}
\begin{cases}
\partial_x\mu^\alpha+ik[\sigma_3,\mu^\alpha]=W^\alpha\,\mu^\alpha\,,\\
\partial_t\mu^\alpha+2ik^2[\sigma_3,\mu^\alpha]=P^\alpha\,\mu^\alpha\,,
\end{cases}
\ee
where
\bea
\label{lax_pair}
&&\sigma_3=\left(\begin{array}{cc}
1 & 0\\
0 & -1
\end{array}\right)~~,~~
W^\alpha(x,t)=\left(\begin{array}{cc}
0& q^\alpha(x,t)\\
g\,\bar{q}^{\alpha}(x,t) & 0
\end{array}\right)\,,\\
&&P^\alpha(x,t,k)=2k\,W^\alpha-i\partial_x W^\alpha\,\sigma_3-i(W^\alpha)^2\,\sigma_3\,.
\eea
At this stage, we have to make several important remarks to clarify the role of the following section and to understand why the system of equations 
\eqref{NLS_edge}-\eqref{BC_edge} is indeed relevant to decribe NLS on a star-graph.

First notice that \eqref{NLS_edge} is a particular case of the general (square) matrix NLS for a $N\times N$ square matrix-valued function $Q(x,t)$
\bea
i\partial_t Q+\partial_x^2 Q-2gQQ^\dagger\,Q=0\,,
\eea
where $Q$ is chosen to be the diagonal matrix with entries $q^1,\dots,q^n$
\bea
\label{Q_diag}
Q(x,t)=\left(\begin{array}{ccc}
q^1(x,t)& &\\
     &\ddots& \\
     & &q^N(x,t)
\end{array}\right)\,.
\eea
This simple observation is the basis of the mapping of \eqref{NLS_edge}-\eqref{BC_edge} to a diagonal matrix IBV problem discussed in the next section. 
In particular, the model we consider \textit{is not}
\bea
i\partial_t q^\alpha+\partial_x^2 q^\alpha-2g\left(\sum_{\beta=1}^N|q^\beta|^2\right)q^\alpha=0\,,~~x,t>0\,,~~\alpha=1,\dots,N\,,
\eea
which would be the vector NLS equation (or general Manakov model \cite{Man}) on the half-line. This would correspond to choosing $Q$ as a row matrix of the form
\bea
Q(x,t)=\left(\begin{array}{ccc}
q^1(x,t)& \dots &q^N(x,t)\\
0 & \dots & 0\\
\vdots     &\ddots& \vdots\\
    0 & \dots &0
\end{array}\right)\,.
\eea
 This model was studied in \cite{CZ1,CZ2} 
for integrable boundary conditions. There, the superscript $\alpha$ refers to 
an internal degree of freedom (light polarization in the historical case $N=2$) and there is a global $U(N)$ symmetry. In the present case, 
the superscript $\alpha$ has the meaning of a discrete spatial coordinate (assuming some embedding of the graph into $\RR^2$ for instance) and the 
global symmetry is only $U(1)^N$.

Second, it is important to realize that the interactions between the edges are mediated only through the 
central vertex via the boundary conditions (\ref{BC_edge}). In our context, the presentation of the problem \eqref{NLS_edge}-\eqref{BC_edge} 
in the standard form of an IBV problem could give the misleading impression that one is in fact always dealing with $N$ disconnected half-lines. This 
would defeat the purpose of this paper which is to deal with equations on a star-graph. In sections \ref{compare} and \ref{general_case}, 
we show that our formulation does indeed include important examples of genuine star-graphs where the half-lines are connected nontrivially. 
Of course, if one was to give oneself the complete
set of boundary data $g_0^\alpha$, $g_1^\alpha$ (either directly or through Dirichlet-to-Neumann maps), then one could simply apply $N$ times the 
unified transform of Fokas to solve the problem on each half-line independently. The point is that in situations of interest for problems on a star-graph, 
only special combinations of $g_0^\alpha$, $g_1^\alpha$ are supposed to be given and hence one has to consider the problem as a whole and cannot split it 
into $N$ disconnected problems. This is reminiscent of the well-known case on the half-line where, generically, one would present the problem as 
\bea
&& i\partial_t q+\partial_x^2 q-2g|q|^2q=0~~,~~x,t>0\,,\\
&& q(x,0)=q_0(x)\,,\\
&& q(0,t)=g_0(t)\,,~~\partial_x q(0,t)=g_1(t)\,,
\eea
but then, one would restrict their attention to the case of Robin boundary condition where only the combination $\eta q(0,t)+\partial_x q(0,t)$ is supposed to be 
given. In other words, only $g_1(t)+\eta g_0(t)\equiv g(t)$ forms the boundary data. The particular case $g(t)=0$ is the well-known integrable Robin boundary condition 
for scalar NLS on the half-line \cite{Sk}. In the same spirit, we will see in the examples of Sections \ref{compare} and \ref{general_case} that only certains combinations of 
$g_0^\alpha$, $g_1^\alpha$ are supposed to be given. Such examples will represent typical situations of a star-graph where the half-lines are 
connected through the vertex.

\section{Mapping the Problem to a Matrix IBV Problem}\label{mapping}

The observation in the previous section that the set of equations \eqref{NLS_edge} is a particular case of a general matrix NLS equation suggests that 
a natural way to deal with the problem on a star-graph is to consider matrix-valued functions of a certain type. Indeed, defining 
\bea
\label{def_WP}
&&\Sigma_3=\left(\begin{array}{cc}
\1_N & 0\\
0 & -\1_N
\end{array}\right)\,,~~
W(x,t)=\left(\begin{array}{cc}
0& Q(x,t)\\
g\,\overline{Q}(x,t) & 0
\end{array}\right)\,,\\
\label{def_WP2}
&&P(x,t,k)=2k\,W-i\partial_x W\,\Sigma_3-iW^2\,\Sigma_3\,.
\eea
with $Q$ given by \eqref{Q_diag}, then the set of equations \eqref{NLS_edge} is the compatibility condition of the following auxiliary problem
\bea
\begin{cases}
\label{Lax_pair1}
\partial_x \mu+ik[\Sigma_3,\mu]=W\,\mu\,,\\
\partial_t \mu+2ik^2[\Sigma_3,\mu]=P\,\mu\,,
\end{cases}
\eea
where all the objects are now $2N\times 2N$ matrices. So let us denote by $\cM_p$ the algebra of $p\times p$ matrices over $\CC$ and $\{E_{ij}\}_{i,j=1}^p$ 
its canonical basis. When dealing with different values 
of $p$ in a single expression, it is understood that the basis matrices $E_{ij}$ have the appropriate size given by the range of their indices.
Given $M\in\cM_{2N}$, define
\bea
M_d&=&\sum_{k,\ell=1}^2\sum_{i=1}^N M_{i+(k-1)N,i+(\ell-1)N}E_{k\ell}\otimes E_{ii}\,,\\
M_a&=&\sum_{k,\ell=1}^2\sum_{1\le i\neq j\le N} M_{i+(k-1)N,j+(\ell-1)N}E_{k\ell}\otimes E_{ij}\,.
\eea
We denote $\cM_d=\{M_d, M\in\cM_{2N}\}$ and $\cM_a=\{M_a, M\in\cM_{2N}\}$ the corresponding sets.
A matrix in $\cM_d$ looks like $\left(\begin{array}{c|c}
M_{1} &M_{2} \\
    \hline
M_{3}&M_{4} \\
\end{array}\right)$
where each $M_j$ is an $N\times N$ diagonal matrix. The point is that $\Sigma_3$, $W$ and $P$ in \eqref{Lax_pair1} are all $\cM_d$-valued functions. 
To implement our formalism, detailed in the next section, the crucial fact is that the fundamental solutions of \eqref{Lax_pair1}, properly normalised, 
will also lie in $\cM_d$ for all values of $x,t,k$ where they are defined. This fact, together with the following simple lemma, will then ensure that all 
the objects appearing in the ISM of a system on a star-graph will also lie in $\cM_d$. As a consequence, the original IBV problem \eqref{NLS_edge}-\eqref{BC_edge} 
representing NLS on a star-graph will be solvable by implementing a generalization of Fokas method to the case of $\cM_d$-valued functions.
\begin{lemma}
\label{lemma1}
$\cM_d$ and $\cM_a$ are vector subspaces of $\cM_{2N}$ and the direct sum decomposition $\cM_{2N}=\cM_d\oplus \cM_a$ holds.
Moreover, $\cM_d$ is a subalgebra of $\cM_{2N}$ which is isomorphic to the direct product $\displaystyle\prod_{i=1}^N\cM_2$
as algebras. 
\end{lemma}
The last point is easily seen using the following isomorphism
\bea
\begin{array}{lrll}
\theta:&\displaystyle\prod_{i=1}^N\cM_2&\to&\cM_d\\
&(M^1,\dots,M^N)&\mapsto& M=\displaystyle\sum_{\alpha=1}^N M^\alpha\otimes E_{\alpha\alpha}
\end{array}\,,
\eea
where the algebra structure of $\prod_{i=1}^N\cM_2$ is defined by the pointwise operations.

In the case $N=1$, the study of the analytic properties of the solutions to Eqs. (\ref{Lax_pair1}) treated simultaneously 
forms the basis of the unified method developed by Fokas for IBV problems \cite{Fokas}. We refer to the $N=1$ case 
as the \textit{scalar} case. For our purposes, the following theorem is a key result of this paper. As explained above, 
it shows that analyzing the problem on a star-graph 
is the same as analyzing a matrix IBV problem in $\cM_d$. 
\begin{theorem}
\label{theorem_mapping}
Let $(x_0,t_0)\in\RR^+\times\RR^+$ and let $\mu$ be the fundamental solution of (\ref{Lax_pair1}) normalised by $\mu(x_0,t_0,k)=\1_{2N}$. 
Then $\mu(x,t,k)\in\cM_d$ wherever it is defined.
\end{theorem}
\prf
Denote $\phi(x,t,k)=kx+2k^2t$. Eqs. (\ref{Lax_pair1}) are equivalent to the equation
\bea
\label{exact_form}
d\left(e^{i\phi\Sigma_3}\mu\, e^{-i\phi\Sigma_3} \right)=e^{i\phi\Sigma_3}\left(W\,\mu\,dx+P\,\mu\,dt\right)e^{-i\phi\Sigma_3}\,,
\eea
ensuring that the right-hand side is an exact $1$-form. Fix $(x_0,t_0)\in\RR^+\times\RR^+$ and define the solution, for $x>0$ and 
$t>0$,
\be
\label{Volterra}
\mu(x,t,k)=\mu_0(k)+\int_{(x_0,t_0)}^{(x,t)}e^{i\phi(y-x,\tau-t,k)\Sigma_3}\left(W(y,\tau)\mu(y,\tau,k)dy+P(y,\tau,k)\mu(y,\tau,k)d\tau\right)
e^{-i\phi(y-x,\tau-t,k)\Sigma_3}
\ee
Using the linearity of the Volterra integral equation (\ref{Volterra}) and Lemma \ref{lemma1} to project it on $\cM_d$ and $\cM_a$, 
we obtain that both 
$\mu_d$ and $\mu_a$ satisfy (\ref{Volterra}). To formulate the ISM, one uses fundamental solutions of 
(\ref{Lax_pair1}) defined by $\mu_0(k)=\1_{2N}$. Therefore, for such a solution $(\mu_{0}(k))_a=0$. By uniqueness of the solution 
of (\ref{Volterra}), we deduce that $\mu_a=0$ identically. Hence, any fundamental solution of (\ref{Lax_pair1}) 
is an $\cM_d$-valued function, which concludes the proof. Note that all the ingredients of the ISM being derived from fundamental solutions 
by algebraic operations and $\cM_d$ being an algebra, we deduce that ISM for the problem 
(\ref{NLS_edge})-(\ref{BC_edge}) can be entirely formulated in $\cM_d$.
\finprf

\textit{Remark $1$:} The values of $k\in\CC$ for which a fundamental solution is defined or has given analytic properties depend on $(x_0,t_0)$. This 
is what we mean in the previous theorem by "wherever it is defined". This will become clear below when we apply our setup to the case of NLS and specify 
three values of $(x_0,t_0)$.

\textit{Remark $2$:} In practice, the theorem ensures that the fundamental solutions of Eqs. (\ref{Lax_pair1}) 
can be split into $N$ $2\times 2$ matrices whenever this is more convenient than their form in $\cM_d$. One simply uses the inverse of the isomorphism $\theta$.
This will be used in the proof of Theorem \ref{reconstruction_theorem}. We also get that all spectral functions $\{a(k),b(k)\}$ and $\{A(k),B(k)\}$ 
appearing in the ISM on a star-graph are $N\times N$ diagonal matrices.

\section{Inverse Scattering Method on a Star Graph}\label{ISM}

Equipped with Theorem \ref{theorem_mapping}, we formulate the ISM on a star graph by extending the unified method of Fokas to matrices in $\cM_d$.
For full details on the method in the scalar case, we refer to the book \cite{Fokas}. In particular, in the following, we will heavily rely on 
Chapter $16$ of \cite{Fokas}. The approach is of analysis/synthesis nature. Under the assumption that $Q(x,t)$ exists, 
the analysis part allows one to introduce the relevant scattering data, the so-called 
\textit{global relation} and to formulate an appropriate Riemann-Hilbert problem which is at the basis of the inverse part of the ISM for IBV problems. 
Equipped with all this, the synthesis part consists in formulating the direct and inverse parts of the ISM, assuming that the global relation holds, 
and then check that the obtained solution $Q(x,t)$ indeed 
satisfies the PDE together with the initial and boundary conditions. Here, we present directly the main results in $\cM_d$ and 
focus on the synthesis part. We give the key steps of the analysis part, pointing out the main differences with the scalar case.

\subsection{Spectral analysis}\label{spectral_analysis}

In Fokas's method for an IBV problem, one needs to define three fundamental solutions $\mu_j(x,t,k)$ as in \eqref{Volterra}, 
corresponding to the three points $(x_0,t_0)=(0,T)$ for $j=1$, $(x_0,t_0)=(0,0)$ for $j=2$ and $(x_0,t_0)=(\infty,t)$ for $j=3$. 
Any pair of these solutions is related by a matrix 
independent of $x,t$. So one defines the two scattering matrices $S(k)$ and $T(k)$ for $k\in\RR$ by
\bea
\label{def1}
&&\mu_3(x,t,k)=\mu_2(x,t,k)e^{-i\phi(x,t,k)\Sigma_3}\,S(k)\,e^{i\phi(x,t,k)\Sigma_3}\,,\\
\label{def2}
&&\mu_1(x,t,k)=\mu_2(x,t,k)e^{-i\phi(x,t,k)\Sigma_3}\,T(k)\,e^{i\phi(x,t,k)\Sigma_3}\,.
\eea
Using the symmetry of the potential $W$
\be
W^\dagger(x,t)=\left(\begin{array}{cc}
g\1_N & 0\\
0 & \1_N
\end{array}\right)W(x,t)\left(\begin{array}{cc}
\frac{1}{g}\1_N & 0\\
0 & \1_N
\end{array}\right),
\ee 
the fact that $\det \mu_j(x,t,k)=1$, $j=1,2,3$ and that our matrices lie in $\cM_d$, we obtain that $S$ and $T$ have the general form
\bea
\label{def_S_T}
S(k)=\left(\begin{array}{cc}
 \overline{a(\bar{k})}& b(k)\\
g\,\overline{b(\bar{k})}& a(k)
\end{array}\right)\,,~~
T(k)=\left(\begin{array}{cc}
 \overline{A(\bar{k})}& B(k)\\
g\,\overline{B(\bar{k})}& A(k)
\end{array}\right),
\eea
where the scattering coefficients $a(k),b(k),A(k),B(k)$ are $N\times N$ diagonal matrices. 
The entries of the matrices $a(k),b(k),A(k),B(k)$ extend to functions on the complex plane with certain analytic properties. 
From the Volterra integral representation \eqref{Volterra}, one deduces that the entries of $a(k)$ and $b(k)$ are analytic for 
$\arg k\in(0,\pi)$ and continuous and bounded for $\arg k\in[0,\pi]$. The entries of $A(k)$ and $B(k)$ are entire functions that are bounded 
for $\arg k\in[0,\frac{\pi}{2}]\cup[\pi,\frac{3\pi}{2}]$ (in the case of finite $T$).
For convenience, we denote by $D_j$ the $j$-th quadrant of the complex plane defined by
\be
D_j=\{z\in\CC,\arg z\in((j-1)\frac{\pi}{2},j\frac{\pi}{2})\}~~,~~j=1,2,3,4\,.
\ee
An important role is played by the zeros of the entries of $a(k)$ and of the matrix $d(k)=a(k)\,\overline{A(\bar{k})}-
g\,b(k)\,\overline{B(\bar{k})}$. Following \cite{Fokas}, we make some assumptions on these zeros. This is where the present matrix case requires extra 
care compared to the scalar case. Hence, we assume
\begin{enumerate}
\item[(C1)] For each $\alpha\in\{1,\dots,N\}$, $a_\alpha(k)$ has $K^\alpha$ simple zeros $\{\kappa_j^\alpha\}_{j=1}^{K^\alpha}$ 
with $K^\alpha=K_1^\alpha+K_2^\alpha$,
$\kappa_j^\alpha\in {D_1}$, $j=1,\dots,K_1^\alpha$; $\kappa_j^\alpha\in {D_2}$, $j=K_1^\alpha+1,\dots,K^\alpha$.

\item[(C2)] For each $\alpha\in\{1,\dots,N\}$, $d_\alpha(k)$ has $\Lambda^\alpha$ simple zeros $\{\lambda_j^\alpha\}_{j=1}^{\Lambda^\alpha}$ with 
$\lambda_j^\alpha\in {D_2}$, $j=1,\dots,\Lambda^\alpha$. 

\item[(C3)] None of the zeros of $a_\alpha(k)$ {in $D_2$} coincides with a zero of $d_\alpha(k)$. 

\end{enumerate}
In general, one can have coinciding zeros for different $\alpha$ and $\beta$ \ie $\kappa_j^\alpha=\kappa_\ell^\beta$ for some $\alpha\neq \beta$ and 
$j,\ell$. Similarly, one could have $\lambda_j^\alpha=\lambda_\ell^\beta$ for some $\alpha\neq \beta$ and $j,\ell$. Finally, one can also have
$\kappa_j^\alpha=\lambda_\ell^\beta$ for some $\alpha\neq \beta$ and $j,\ell$. Such instances correspond in general to the situation where
 the different half-lines are non trivially connected.

Another important ingredient is the global relation. The same derivation as in \cite{Fokas} goes over to our matrix case. 
From the integration of the exact form \eqref{exact_form} around the boundary of the domain
$\{0<x<\infty,0<t<T\}$, one arrives at
\bea
\1_{2N}-T^{-1}(k)\,S(k)-\int_{0}^{\infty}e^{i\phi(x,T,k)\Sigma_3}\,W(x,T)\,\mu_3(x,T,k)\,e^{-i\phi(x,T,k)\Sigma_3}\,dx=0\,.
\eea
The blocks $(11)$ and $(21)$ of this relation hold for $k\in \overline{D}_3\cup \overline{D}_4$ while the blocks $(12)$ and $(22)$ hold for 
$k\in\overline{D}_1\cup \overline{D}_2$. In particular, the block $(12)$ of this relation yields the \textit{global relation}
\bea
a(k)B(k)-b(k)A(k)=e^{4ik^2T}c(T,k)~~,~~k\in\overline{D}_1\cup \overline{D}_2\,,
\eea
where $c(T,k)=\int_{0}^{\infty}e^{2ikx}\,Q(x,T)\,(\mu_3(x,T,k))_{22}\,dx$ is analytic for $k\in D_1\cup D_2$ and of order $\frac{1}{k}$ as $k\to\infty$. 
In the case $T=\infty$, this boils down to 
\bea
\label{global_infty}
a(k)B(k)-b(k)A(k)=0~~,~~k\in\overline{D}_1\,.
\eea

\subsection{Synthesis: direct and inverse transforms of ISM}

\subsubsection{Direct part}\label{direct_part}

Consider the initial-boundary data $q_0^j$, $g_0^j$, $g_1^j$, $j=1,\dots,N$ from \eqref{IC_edge}-\eqref{BC_edge} and assume 
that $q_0^j$ is of Schwarz type and $g_0^j$, $g_1^j$ are smooth functions for $0<t<T$, $j=1,\dots,N$. Denote 
$Q_0(x)=diag\left(q_0^1(x),\dots,q_0^N(x)\right)$ and $H_\ell(t)=diag\left(g_\ell^1(t),\dots,g_\ell^N(t)\right)$, $\ell=0,1$. 
Also, let 
\bea
W_0(x)=\left(\begin{array}{cc}
0 &Q_0(x)\\
g\overline{Q}_0(x) & 0
\end{array}\right)
\,,~~G_\ell(x)=\left(\begin{array}{cc}
0 &H_\ell(t)\\
g\overline{H}_\ell(t) & 0
\end{array}\right)\,,~~\ell=0,1\,.
\eea
Now, define $\varphi(x,k)$, $\Phi(t,k)$ as the $2N\times N$ matrix-valued functions satisfying
\begin{subequations}
\begin{align}
\label{def_wavefunctions1}
&\partial_x\varphi+2ik\left(\begin{array}{cc}
\1_N &0\\
0 & 0
\end{array}\right)\varphi=W_0\,\varphi~~,~~0<x<\infty~~,~~k\in\overline{D}_1\cup\overline{D}_2\,,\\
\label{def_wavefunctions2}
&\partial_t\Phi+4ik^2\left(\begin{array}{cc}
\1_N &0\\
0 & 0
\end{array}\right)\Phi=(2kG_0-iG_1\Sigma_3-iG_0^2\Sigma_3)\,\Phi\,,~~0<t<T\,,~~k\in\overline{D}_1\cup\overline{D}_3\,,\\
\label{def_wavefunctions3}
&\displaystyle\lim_{x\to\infty}\varphi(x,k)=\left(\begin{array}{c}
0\\
\1_N
\end{array}\right)\,,~~\displaystyle\lim_{t\to T}\Phi(t,k)=\left(\begin{array}{c}
0\\
\1_N
\end{array}\right)\,.
\end{align}
\end{subequations}
The scattering coefficients are defined by 
\bea
\label{def_ab}
\left(\begin{array}{c}
b(k)\\
a(k)
\end{array}\right)=\varphi(0,k)~~\text{and}~~\left(\begin{array}{c}
B(k)\\
A(k)
\end{array}\right)=\Phi(0,k)\,.
\eea
Finally, we assume that $Q_0$, $H_0$ and $H_1$ are such that: $Q_0(0)=H_0(0)$, $Q_{0x}(0)=H_1(0)$ and conditions $(C1)-(C3)$ hold as well as 
the global relation.

\subsubsection{Inverse part}\label{inverse_part}

Given the scattering coefficients $a(k),b(k),A(k),B(k)$ together with the zeros as in $(C1)-(C3)$, define the matrix $J$ by
$J(k)=J_\ell$ when $\arg k=\frac{\ell\pi}{2}$, where
\bea
\label{defJ_first}
&&J_1=\left(\begin{array}{cc}
\1_N&0\\
\Gamma(k)\,e^{2i\phi(x,t,k)} & \1_N
\end{array}\right)\,,~~
J_4=\left(\begin{array}{cc}
\1_N&-\gamma(k)\,e^{-2i\phi(x,t,k)}\\
g\bar{\gamma}(k)\,e^{2i\phi(x,t,k)} & \1_N-g|\gamma(k)|^2
\end{array}\right)\,,\\
&&J_3=\left(\begin{array}{cc}
\1_N&-g\overline{\Gamma(\bar{k})}\,e^{-2i\phi(x,t,k)}\\
0 & \1_N
\end{array}\right)\,,~~
J_2=J_3J_4^{-1}J_1\,,
\eea
and
\bea
\label{def_gammas}
&&\gamma(k)=b(k)\,\bar{a}^{-1}(k)\,,k\in\RR\,,~~\Gamma(k)=g\overline{B(\bar{k})}a^{-1}(k)d^{-1}(k)\,,~~k\in\RR^-\cup i\RR^+\,,\\
\label{defJ_last} &&d(k)=a(k)\,\overline{A(\bar{k})}-g\,b(k)\,\overline{B(\bar{k})}\,,~~k\in\RR^-\cup i\RR^+\,.
\eea
Then, define $M(x,t,k)$ as the solution of the following $2N\times 2N$ matrix Riemann-Hilbert problem
\begin{itemize}
\item $M$ is meromorphic for $k\in\CC\setminus\{\RR\cup i\RR\}$;

\item $M_-(x,t,k)=M_+(x,t,k)\,J(x,t,k)\,,~~k\in\RR\cup i\RR$
where $M=M_-$ for $k$ in the second or fourth quadrant, $M=M_+$ for $k$ in the first or third quadrant and $J$ is defined in terms of $a,b,A,B$ as in 
\eqref{defJ_first}-\eqref{defJ_last};

\item $M(x,t,k)=\1_{2N}+O\left(\frac{1}{k}\right)\,,~~k\to\infty$;

\item Dropping the $x,t$ dependence for conciseness and denoting $M(k)=([M]_1(k),[M]_2(k))$ the splitting of $M$ into the first and last $N$ columns,
the following residue conditions hold at the possible zeros of the entries of $a(k)$ and $d(k)$
\bea
\label{res1}
\mathop{Res}_{\kappa_j^\alpha}[M]_1&=&e^{2i\phi(\kappa_j^\alpha)}\,[M]_2(\kappa_j^\alpha)\,b^{-1}(\kappa_j^\alpha)\,\mathop{Res}_{\kappa_j^\alpha}a^{-1}\,,\\
\mathop{Res}_{\bar{\kappa}_j^\alpha}[M]_2
&=&e^{-2i\phi(\bar{\kappa}_j^\alpha)}\,[M]_1(\bar{\kappa}_j^\alpha)\,g^{-1}\bar{b}^{-1}(\kappa_j^\alpha)\,\mathop{Res}_{\bar{\kappa}_j^\alpha}\bar{a}^{-1}\,,\\
\mathop{Res}_{\lambda_j^\alpha}[M]_1&=&
ge^{2i\phi(\lambda_j^\alpha)}\,[M]_2(\lambda_j^\alpha)\,\overline{B(\bar{\lambda}_j^\alpha)}\,a^{-1}_0(\lambda_j^\alpha)\,\mathop{Res}_{\lambda_j^\alpha}d^{-1}\,,\\
\label{res4}
\mathop{Res}_{\bar{\lambda}_j^\alpha}[M]_2&=&
e^{-2i\phi(\bar{\lambda}_j^\alpha)}\,[M]_1(\bar{\lambda}_j^\alpha)\,B(\bar{\lambda}_j^\alpha)\,\bar{a}^{-1}_0(\lambda_j^\alpha)\,\mathop{Res}_
{\bar{\lambda}_j^\alpha}\bar{d}^{-1}\,,
\eea
where $\displaystyle\mathop{Res}_{\kappa_j^\alpha}a^{-1}$ is the diagonal matrix whose only nonzeros entries are for those $\beta$'s such that 
$\kappa_j^\alpha$ is a zero of $a_\beta(k)$, in which case the element reads $\frac{1}{\dot{a}_\beta(\kappa_j^\alpha)}$, and 
similarly for $\displaystyle\mathop{Res}_{\bar{\kappa}_j^\alpha}\bar{a}^{-1}$, $\displaystyle\mathop{Res}_{\lambda_j^\alpha}d^{-1}$ and 
$\displaystyle\mathop{Res}_{\bar{\lambda}_j^\alpha}\bar{d}^{-1}$.

\end{itemize}

The fundamental result of ISM for NLS on a star graph is then the following 
\begin{theorem}
\label{reconstruction_theorem}
$M(x,t,k)$ exists and is unique. Moreover, if we set
\bea
\label{recons}
Q(x,t)=2i\lim_{k\to\infty}(kM(x,t,k))_{12}\,,
\eea
then $Q(x,t)$ solves the NLS equation on a star-graph with initial condition $Q(x,0)=Q_0(x)$ and boundary conditions 
$Q(0,t)=H_0(t)$, $\partial_x Q(0,t)=H_1(t)$. The index "$12$" in \eqref{recons} means that we take the block $(12)$ in the natural decomposition of matrices in 
$\cM_d$.
\end{theorem}
\prf
Using the isomorphism $\theta$, we can map the proof of this theorem to the proof of $N$ copies of the analogous theorem for the scalar case. Indeed, let 
$(M^1,\dots,M^N)$ be the preimage of $M$ by $\theta$ then each $M^\alpha$ is defined as the solution of the $2\times 2$ Riemann-Hilbert problem analogous 
to the one presented above but based on the scattering data $a_\alpha(k)$, $b_\alpha(k)$, $A_\alpha(k)$, $B_\alpha(k)$ corresponding to $q_0^\alpha(x)$, 
$h_j^\alpha(t)$, $j=1,2$. In this case, it is known that $M^\alpha$ exists and is unique \cite{Fokas_CMP}. Now, setting $\displaystyle Q(x,t)=2i\lim_{k\to\infty}(kM(x,t,k))_{12}$ is equivalent to setting 
$\displaystyle q^\alpha(x,t)=2i\lim_{k\to\infty}
(kM^\alpha(x,t,k))_{12}$ for $\alpha=1,\dots,N$. In the last equation, the index $12$ is simply the entry in position $(12)$ in the $2\times 2$ matrix 
$kM^\alpha$. Again, it is a consequence of the results in \cite{Fokas_CMP} that $q^\alpha$ is then solution of NLS on the half-line 
which satisfies the initial condition $q^\alpha(x,0)=q_0^\alpha(x)$ and boundary conditions $q^\alpha(0,t)=h_0^\alpha(t)$, $\partial_x q^\alpha(0,t)=h_1^\alpha(t)$.
This means that $Q(x,t)$ satisfies NLS on a star-graph with initial-boundary data $Q_0,H_0,H_1$.
\finprf
 
\subsection{The global relation}
In general, the success of the whole method relies on the ability to analyse the global relation, that is, to obtain a characterization of the 
spectral functions $A(k)$ and $B(k)$ compatible with $a(k)$, $b(k)$\footnote{We note though that the use of the nonlinear steepest descent 
method in the case of boundary conditions that decay for large $t$ allows to obtain asymptotic results without the explicit analysis of the global relation.},
 the latter being determined by the initial data $Q_0$. 
For instance, given the initial and Dirichlet data, $Q_0(x)$ and 
$H_0(t)$, (or the initial and Neumann data, $Q_0(x)$ and $H_1(t)$), one can ensure that the whole enterprise will be successful by showing that 
the remaining data $H_1(t)$ (or $H_0(t)$) can be consistently chosen so as to ensure that the global relation holds. This is the 
Dirichlet-to-Neumann (or Neumann-to-Dirichlet) map problem which is strongly related to the analysis of the global relation. 
This has been recognized for a long time in the scalar case and two main approaches are available: 
\begin{itemize}
\item Either one can identify boundary conditions for which the whole problem can be avoided and such that the data $\{A(k),B(k)\}$ can 
be eliminated from the reconstruction procedure of $Q(x,t)$. These are the so-called linearizable boundary conditions;
\item Or one can find an explicit formula for the missing data, say $H_1(t)$, in terms of quantities depending only on $Q_0(x)$ and $H_0(t)$ 
(if one deals with the Dirichlet-to-Neumann map), and then obtain the corresponding $A(k)$, $B(k)$.
\end{itemize}
In the next section, we present a detailed study of a special instance of the first possibility by recovering the solution of the scalar initial value 
problem on the full line from our representation as a matrix initial-boundary value problem. 
In the rest of this section, we discuss the second possibility. In particular, thanks to Theorem \ref{theorem_mapping}, we are able to transfer to the 
present case the important result on the Dirichlet-to-Neumann map found in \cite{Fokas_DM_map} in the scalar case. To our knowledge, the result of 
\cite{Fokas_DM_map} is the best one obtained under general assumptions on the initial-boundary data, in the sense that it gives 
an explicit formula for the Neumann data in terms of the initial and Dirichlet data via the solution of coupled nonlinear ODEs. 
In the case of NLS, the latest developments on this problem are presented in \cite{FL1} for choices of Dirichlet and Neumann 
data with a specific, asymptotically periodic, behaviour. They build on the previous paper \cite{FL2} and show how effective the Dirichlet-to-Neumann 
map can be. For our purposes here, we present the star-graph version of the results of \cite{Fokas_DM_map}.
\begin{proposition}\label{solution_H1}
Let $Q_0$ and $H_0$ be given such that the entries of $Q_0$ are Schwarz functions on $\RR^+$, the entries of $H_0(t)$ are smooth functions
for $0<t<T$,
and $Q_0(0)=H_0(0)$. Assume that the entries of $a(k)$ have no zeros in $D_1$. Then, the Neumann data has the following explicit form 
\bea
\label{form_H1}
H_1(t)&=& \frac{2}{\pi}H_0(t)\int_{\partial D_1}\left(\Phi_2(t,k)-\Phi_2(t,-k)\right)dk\nonumber\\
&&+\frac{2i}{\pi}\int_{\partial D_1}\left[k \left(\Phi_1(t,k)-\Phi_1(t,-k)\right)+iH_0(t)\right]dk\nonumber\\
&&+\frac{4i}{\pi}\int_{\partial D_1}k e^{-4ik^2t}b(k)a^{-1}(k)\overline{\Phi_2(t,\bar{k})}\,dk
\eea
where $a(k),b(k)$ are defined from $Q_0$ as in \eqref{def_ab} and the contour $\partial D_1$
is the boundary of the quadrant $D_1$ oriented from $i\infty$ to $\infty$ via $0$. Here, $\Phi_1(t,k)$, $\Phi_2(t,k)$ are the two $N\times N$ diagonal matrices 
forming $\Phi(t,k)$ satisfying \eqref{def_wavefunctions2}, in which $H_1(t)$ (appearing in $G_1(t)$) should be replaced by the above expression, and
with initial condition
\be
\Phi(0,k)=\left(\begin{array}{c}
0\\
\1_N
\end{array}\right)\,.
\ee
In this case, the scattering coefficients $A(k)$ and $B(k)$ 
can be deduced from the formulas
\be
A(k)=\overline{\Phi_2(T,\bar{k})}\,,~~B(k)=-\Phi_1(T,k)e^{4ik^2T}\,.
\ee
\end{proposition}
Note that, using Theorem \ref{theorem_mapping}, this proposition can be readily obtained by applying formula $(1.16)$ of \cite{Fokas_DM_map} 
to each component $g_1^j(t)$, $j=1,\dots,N$ of $H_1(t)$. As a matter of fact, the whole argument of \cite{Fokas_DM_map} can be transferred 
component by component to the present context. However, for the reader's convenience, we present a proof in Appendix \ref{AppA}.\\
\textit{Remark $3$:}
The assumption on the entries of $a(k)$ is for simplicity only and the case with a finite number of simple zeros can be easily obtained by using the 
residue theorem. \\
\textit{Remark $4$:} The usefulness of this result relies on the ability to determine $\Phi_1(t,k)$ and $\Phi_2(t,k)$. 
They satisfy the differential equation \eqref{def_wavefunctions2} 
which becomes nonlinear in view of the replacement of $H_1(t)$ in $P(0,t,k)$ by its explicit expression \eqref{form_H1}. Equivalently, 
$\Phi_1(t,k)$ and $\Phi_2(t,k)$ satisfy coupled nonlinear Volterra integral equations.
 In the scalar case, the existence of a solution is discussed in \cite{Fokas_DM_map}. Moreover, it is shown in 
\cite{FL2} that $\Phi_1(t,k)$ and $\Phi_2(t,k)$ can be found efficiently by using a perturbative scheme which yields exact solutions to all orders of the scheme. 
Obvioulsy, once again the same is true for the present star-graph case by simply applying the method to each component of our diagonal matrices.

\section{Comparison with previous results}\label{compare}

Obviously, the scalar case $N=1$ boils down to NLS on the half-line which has been the object of numerous studies \cite{Sk,T,BT,Fokas}. 
For $N\ge 2$, the apparent simplicity of the proof of the main theorem in Section \ref{mapping} when one uses the map $\theta$ is both beautiful and potentially misleading. 
One may erroneously infer that we are simply dealing with $N$ disconnected copies of the half-line problem. It is the object of this section 
to show that this is not so and that our approach actually unifies and encompasses all previous studies (known to the author) 
of the NLS equation on more complicated supports than the full line.

 \subsection{Case $N=2$: problem on the line with a defect/impurity}

 \subsubsection{Recovering the problem on the line}
 
 The simplest way to check that our formalism does describe connected half-lines is to show how it reproduces the problem on the full line.\footnote{We are grateful to 
 N. Cramp\'e for this useful observation.} The latter can be seen 
 as the problem on two half-lines connected in such a way that there is no reflection and trivial transmission. In fact, the crux of the matter can already been seen 
 for the linear case \ie when the coupling constant $g=0$ in NLS. The point is that the boundary conditions encoded in the functions $H_0=diag(g_0^1,g_0^2)$ and 
 $H_1=diag(g_1^1,g_1^2)$ must 
 "disappear" from the reconstruction formula for the function on the full line and only the initial condition must play a role. 
 
 \paragraph{Linear case.}
  It is very instructive to 
 look at the details in the linear case for $N=2$ first. From the half-line problem 
 \bea
 \label{full_half}
 &&i\partial_t Q(x,t)+\partial_x^2 Q(x,t)=0\,,~~x,t>0\,,\\
 &&Q(x,0)=Q_0(x)\,,~~x\ge 0\,,\\
 &&Q(0,t)=H_0(t)\,,~~\partial_xQ(0,t)=H_1(t)\,,~~t\ge 0\,,
 \eea
 we want to solve the full line problem
 \bea
 \label{full_full}
&& i\partial_t u(x,t)+\partial_x^2 u(x,t)=0\,,~~t>0,~x\in\RR\,,\\
&& u(x,0)=u_0(x)\,,~~x\in\RR\,.
 \eea
 This goes as follows. We use the two functions $q_j(x,t)$, $j=1,2$ contained in $Q(x,t)$ and defined on each half-line to 
 form 
 \be
 u(x,t)=\theta(x)q^1(x,t)+\theta(-x)q^2(-x,t)\,.
 \ee 
 The "transparent" boundary conditions are obtained for $g_0^1(t)=g_0^2(t)\equiv g_0(t)$ and 
 $g_1^1(t)=-g_1^2(t)\equiv g_1(t)$ where $g_j^\alpha$ are the boundary data in \eqref{BC_edge}. This corresponds to $H_0(t)=g_0(t)\1_2$ and $H_1(t)=g_1(t)\sigma_3$.
  To compare \eqref{full_half} and \eqref{full_full} more efficiently, let us 
 define 
 \bea
 \label{def_Qline}
 Q^{line}(x,t)=\sigma_3\left(\theta(x)Q(x,t)+\theta(-x)\sigma Q(-x,t)\sigma\right)=
 \left(\begin{array}{cc}
 u(x,t) & 0\\
 0 & -u(-x,t)
 \end{array}\right)\,,
 \eea
 where 
 \be
 \sigma=\left(\begin{array}{cc}
0 & 1\\
1 & 0
\end{array}\right)
 \ee
 Obviously, one simply extracts the $(11)$ entry of $Q^{line}$ to get $u$. So, for convenience, we perform the analysis directly on $Q^{line}$. 
 The prefactor $\sigma_3$ is purely conventional here but will turn out to be useful
 when we go over to the nonlinear case.
The usual Fourier transform method applied to $Q^{line}$ yields
\be
 Q^{line}(x,t)=\frac{1}{2\pi}\int_{-\infty}^\infty \widehat{Q}^{line}(k,t)\,e^{-ikx}\,dk\,,
\ee 
where
\bea
\label{Fourier_line}
\widehat{Q}^{line}(k,t)&=&\int_{-\infty}^\infty Q^{line}(x,t)\,e^{ikx}\,dx\nonumber\\
&=&\sigma_3\int_0^\infty\left(Q(x,t)e^{ikx}+\sigma Q(x,t)\sigma e^{-ikx}\right)\,dx\nonumber\\
&=&\sigma_3\left(\widehat{Q}(k,t)+\sigma\widehat{Q}(-k,t)\sigma\right)\,,
\eea
where $\widehat{Q}(k,t)$ is the (half) Fourier transform of $Q(x,t)$. Now the unified method applied to the problem \eqref{full_half} provides the solution 
for $Q(x,t)$ by deriving the global relation 
\be
\label{linear_GR}
\widehat{Q}(k,t)=e^{-ik^2t}\widehat{Q}(k,0)+e^{-ik^2t}\int_0^te^{ik^2\tau}\left(iH_1(\tau)+kH_0(\tau)\right)\,d\tau
\ee
and then the inverse tranform
\be
Q(x,t)=\frac{1}{2\pi}\int_{-\infty}^\infty \widehat{Q}(k,t)\,e^{-ikx}\,dk\,.
\ee
 The key is the global relation. Here, since $H_0(t)=g_0(t)\1_2$ and $H_1(t)=g_1(t)\sigma_3$, we find from \eqref{linear_GR} that 
 \be
 \widehat{Q}(k,t)+\sigma\widehat{Q}(-k,t)\sigma=e^{-ik^2t}\left(\widehat{Q}(k,0)+\widehat{Q}(-k,0)\right)\,.
 \ee
 Therefore, $H_0$ and $H_1$ have been eliminated from the reconstruction formula and we find 
 \be
 Q^{line}(x,t)=\frac{1}{2\pi}\int_{-\infty}^\infty \widehat{Q}^{line}(k,0)\,e^{-ikx-ik^2t}\,dk\,,
\ee 
as we should, where $\widehat{Q}^{line}(k,0)$ is determined only from the initial condition $Q^{line}(x,0)$. 

\paragraph{NLS case.}
The nonlinear case is 
technically more difficult but the main steps follow the same principle. It provides an important illustration of the role of the 
global relation. We discuss the case $T=\infty$ here for conciseness.
From the half-line problem,
\bea
 \label{half_NLS}
 &&i\partial_t Q(x,t)+\partial_x^2 Q(x,t)=2g\overline{Q}Q^2(x,t)\,,~~x,t>0\nonumber\\
 &&Q(x,0)=Q_0(x)\,,~~x\ge 0\,,\nonumber\\
&& Q(0,t)=g_0(t)\1_2\,,~~\partial_xQ(0,t)=g_1(t)\sigma_3\,,~~t\ge 0\,,
 \eea
 we want to solve the full-line problem
 \bea
 \label{full_NLS}
 && i\partial_t u(x,t)+\partial_x^2 u(x,t)=2g|u|^2u(x,t)\,,~~x\in\RR~,~~t>0\,,\nonumber\\
 &&u(x,0)=u_0(x)\,,~~x\in\RR\,.
 \eea
 We define  $Q^{line}$ as in \eqref{def_Qline} and from it $W^{line}$ and $P^{line}$ as in \eqref{def_WP}, \eqref{def_WP2}. The goal is to show that 
 $Q^{line}(x,t)$ (and hence $u(x,t)$) only depends on $u_0(x)=\theta(x)q_0^1(x)+\theta(-x)q_0^2(-x)$.  
 As is well-known for the standard ISM, the solution for $Q^{line}$ is obtained 
 through the spectral analysis of the two fundamental solutions $\mu_\pm^{line}(x,t,k)$ of 
 \bea
\begin{cases}
\label{Lax_line}
\partial_x \mu^{line}+ik[\Sigma_3,\mu^{line}]=W^{line}\,\mu^{line}\,,\\
\partial_t \mu^{line}+2ik^2[\Sigma_3,\mu^{line}]=P^{line}\,\mu^{line}\,,
\end{cases}
\eea
normalised as
\be
\lim_{x\to\pm\infty}e^{ikx+2ik^2t}\mu_\pm^{line}(x,t,k)e^{-ikx-2ik^2t}=\1_4\,,~~k\in\RR\,.
\ee
In particular, the scattering matrix on the line
\be
S^{line}(k)\equiv\left(
\begin{array}{cc}
\overline{a^{line}(\bar{k})} & b^{line}(k)\\
g\,\overline{b^{line}(\bar{k})} & a^{line}(k)
\end{array}\right)\,,
\ee
satisfies
\be
\mu_+^{line}(x,t,k)=\mu_-^{line}(x,t,k)\,e^{-ikx-2ik^2t}\,S^{line}(k)\,e^{ikx+2ik^2t}\,,~~k\in\RR\,.
\ee
It is also known from the usual ISM that the scattering coefficients $a^{line}(k)$ and $b^{line}(k)$, and their properties in the complex plane,
are completely determined by the initial condition $u_0(x)$. Therefore, it is sufficient to prove that $Q^{line}$ depends only on those coefficients.
For the problem on the half-line, as we discussed in detail in Section \ref{ISM}, 
one must consider 
three fundamental solutions $\mu_j(x,t,k)$ which allow one to define the initial and boundary scattering matrices $S(k)$ and $T(k)$ as in \eqref{def1} and 
\eqref{def2}. In particular, we know that $S(k)=\mu_3(0,0,k)$.
We first need the following lemma on the symmetries of the scattering data.
\begin{lemma}
In the case of transparent boundary conditions, the scattering matrices $S^{line}(k)$, $S(k)$ and $T(k)$ satisfy the following relations
\bea
\label{relation_SSline}
&&I_3\,S(k)\,I_3=\Sigma I_3\,S(-k)\,I_3\Sigma\,S^{line}(k)\,,\\
\label{relation_T}&& T(k)=\Sigma_3\,\Sigma\,T(-k)\,\Sigma\,\Sigma_3\,,
\eea
where $\Sigma=\1_2\otimes \sigma$ and 
\be
I_3=\left(
\begin{array}{cc}
\sigma_3 & 0\\
0 & \1_2
\end{array}\right)\,.
\ee
When projected on the entries of the matrices, these relations extend to the domain of analyticity of the coefficients.
\end{lemma}
\prf
The first observation is that, if $\mu^{line}(x,t,k)$ is a solution of \eqref{Lax_line}, then so is $\Sigma\,\mu^{line}(-x,t,-k)\,\Sigma$. 
This is a consequence of the symmetry\footnote{This is where the prefactor $\sigma_3$ in the definition of $Q^{line}$ is 
important.}
\be
\Sigma\,W^{line}(-x,t)\,\Sigma=-W^{line}(x,t)\,.
\ee
In particular, by uniqueness of normalized solutions, we obtain that 
\be
\mu_-^{line}(x,t,k)=\Sigma\,\mu_+^{line}(-x,t,-k)\,\Sigma\,,
\ee
and as a consequence,  
\be
\label{rel_Sline}
\mu_+^{line}(x,t,k)=\Sigma\,\mu_+^{line}(-x,t,-k)\,\Sigma\,e^{-ikx-2ik^2t}\,S^{line}(k)\,e^{ikx+2ik^2t}\,.
\ee
 The second observation is that, for $x>0$, $I_3\,\mu_3(x,t,k)\,I_3$ and 
$\mu_+^{line}(x,t,k)$ satisfy the same equations and have the same normalization as $x\to\infty$. Hence, for $x>0$,
\be
I_3\,\mu_3(x,t,k)\,I_3=\mu_+^{line}(x,t,k)\,,~~I_3=\left(\begin{array}{cc}
\sigma_3 & 0\\
0 & \1_2
\end{array}\right)\,.
\ee
Taking $t=0$ and the limit as $x$ tends to $0$ of this relation and inserting in \eqref{rel_Sline}, we obtain \eqref{relation_SSline}.
Finally, we note that both $\mu_1(0,t,k)$ and $\mu_2(0,t,k)$ are solutions of 
\be
\partial_t \mu+2ik^2[\Sigma_3,\mu]=P(0,t,k)\,\mu\,,
\ee
where 
\be
P(0,t,k)=\left(\begin{array}{cc}
-igH_0(t)\overline{H}_0(t) & 2kH_0(t)+iH_1(t)\\
g(2k\overline{H}_0(t)-i\overline{H}_1(t)) & igH_0(t)\overline{H}_0(t)
\end{array}\right)\,,
\ee
 contains the boundary data $H_0(t)$ and $H_1(t)$. Therefore, given that 
 \be
H_1(t)+\sigma\,H_1(t)\,\sigma=0\,,~~H_0(t)-\sigma\,H_0(t)\,\sigma=0\,,
 \ee
 we see that
 \be
 P(0,t,k)=\Sigma_3\,\Sigma\,P(0,t,-k)\,\Sigma\,\Sigma_3.
 \ee
 This implies the same symmetry on $\mu_1(0,t,k)$ and $\mu_2(0,t,k)$ which, in turn, yields \eqref{relation_T}.
\finprf

Eq \eqref{relation_SSline} is the nonlinear analog of \eqref{Fourier_line} (at $t=0$) and plays the same crucial role. 
Indeed, in the limit where the coupling $g$ tends to $0$, one has 
\be
S(k)\to\left(\begin{array}{cc}
\1_2 & \widehat{Q}(k,0)\\
0 & \1_2
\end{array}\right)\,,~~S^{line}(k)\to\left(\begin{array}{cc}
\1_2 & \widehat{Q}^{line}(k,0)\\
0 & \1_2
\end{array}\right)\,,
\ee
and the block $(12)$ of \eqref{relation_SSline} yields exactly \eqref{Fourier_line}. 

Combining the lemma with the global relation, we are now ready to prove the following
\begin{proposition}
Given the initial data $q^j_0(x)$ of Schwarz type, $x\ge 0$, $j=1,2$, there is a unique solution $Q^{line}(x,t)$ of 
the nonlinear Schr\"odinger on the line with, defined as in \eqref{def_Qline}, with initial value $u(x,0)=u_0(x)=\theta(x)q^1(x,0)+\theta(-x)q^2(-x,0)$.
\end{proposition}
\prf
In view of \eqref{def_Qline} and of the reconstruction formula \eqref{recons} for $Q(x,t)$, it is sufficient to prove that the solution 
$M(x,t,k)$ of the Riemann-Hilbert problem described in Section \ref{inverse_part} depends only on the spectral functions 
$a^{line}(k)$ and $b^{line}(k)$ which are defined in terms of $u_0$ only. So we have 
to eliminate the matrix data $a(k),b(k),A(k),B(k)$, coming from the boundary problem, in favour of $a^{line}(k)$ and $b^{line}(k)$ only.
This Riemann-Hilbert problem depends on the two quantities $\gamma(k)$ and $\Gamma(k)$ defined in \eqref{def_gammas}. 
From \eqref{relation_SSline}, we extract
\bea
\label{line_1}
a(k)&=&g\sigma\overline{b(-\bar{k})}\sigma_3\sigma b^{line}(k)+\sigma a(-k)\sigma a^{line}(k)\,,\\
\label{line_2}
\sigma_3b(k)&=&\sigma\overline{a(-\bar{k})}\sigma b^{line}(k)+\sigma\sigma_3 b(-k)\sigma a^{line}(k)\,.
\eea
We use \eqref{line_1} into itself to eliminate $a(-k)$ and \eqref{line_2} to express $\overline{b(-\bar{k})}\sigma_3$ and insert into \eqref{line_1}. The net 
result is 
\be
\label{net}
a(k)F^{line}(k)=\overline{b(\bar{k})}G^{line}(k)\,,
\ee 
where $F^{line}(k)$ and $G^{line}(k)$ only depend on $a^{line}(k)$ and $b^{line}(k)$ and are given by
\bea
F^{line}(k)&=&\1_2-g\sigma \overline{b^{line}(-\bar{k})}\sigma b^{line}(k)-\sigma a^{line}(-k)\sigma a^{line}(k)\,,\\
G^{line}(k)&=&g\sigma_3\left[\sigma \overline{a^{line}(-\bar{k})}\sigma b^{line}(k)+\sigma b^{line}(-k)\sigma a^{line}(k) \right]\,.
\eea
Given the definition of $\gamma(k)$, this immediately implies that it depends only on $a^{line}(k)$ and $b^{line}(k)$. To conclude for $\Gamma(k)$, we 
need a bit more work. Eq \eqref{relation_T} yields the symmetry relations
\be
\label{sym_AB}
A(-k)=\sigma A(k)\sigma\,,~~B(-k)=-\sigma B(k)\sigma\,.
\ee
Using these in the definition of $d(k)$ in \eqref{defJ_last} and invoking the global relation \eqref{global_infty}, we obtain the system
\be
\begin{cases}
a(k)B(k)-b(k)A(k)=0\,,\\
\overline{a(-\bar{k})}\sigma A(k)\sigma+g\overline{b(-\bar{k})}\sigma B(k)\sigma=\overline{d(-\bar{k})}\,,
\end{cases}
\ee
for $A(k)$ and $B(k)$. Solving and inserting in the definition of $\Gamma(k)$, we obtain the expression
\be
\Gamma(k)=g\overline{b(\bar{k})}\,\overline{a^{-1}(\bar{k})}\,a^{-1}(k)\left(a(k)-gb(k)\overline{b(\bar{k})}\,\overline{a^{-1}(\bar{k})} \right)^{-1}\,.
\ee
Using the following consequence of \eqref{sym_AB} and the global relation,
\be
b(-k)a^{-1}(-k)=-\sigma b(k)a^{-1}(k)\sigma\,,
\ee
we can bring $\Gamma(k)$ into the convenient form
\be
\Gamma(k)=g\left(\overline{b(\bar{k})} a^{-1}(k)\right)\left(\overline{a^{-1}(\bar{k})}\sigma \overline{a(-\bar{k})}\sigma \right)
\left(a(k)\sigma \overline{a(-\bar{k})}\sigma+gb(k)\sigma \overline{b(-\bar{k})}\sigma \right)\,.
\ee
The first bracket is dealt with using \eqref{net}. The second bracket reads
\be
\overline{a^{-1}(\bar{k})}\sigma \overline{a(-\bar{k})}\sigma=g\overline{b(\bar{k})} a^{-1}(k)\sigma_3\sigma b^{line}(-k)\sigma+\sigma a^{line}(-k)\sigma\,,
\ee
upon using \eqref{line_1}, and hence only depends on $a^{line}(k)$ and $b^{line}(k)$ on account of \eqref{net} again. Finally, the third bracket is seen to be 
simply $a^{line}(k)$ by rewriting \eqref{relation_SSline} as
\be
(S^{line}(k))^{-1}=I_3\,S^{-1}(k)\,I_3\Sigma I_3\,S(-k)\,I_3\Sigma\,,
\ee
and remembering that $\det S(k)=\det S^{line}(k)=1$. 

Therefore, we have shown that the Riemann-Hilbert defining $M(x,t,k)$, and hence $Q(x,t)$, is completely (and only) determined by $a^{line}(k)$ and $b^{line}(k)$, 
as required.
\finprf

\subsubsection{NLS with a $\delta$ potential/impurity}

The $\delta$ potential is the most famous member of a family of singular point potentials (see e.g. \cite{Alb}) and is characterized by one real parameter 
$\eta$. In our setting, it corresponds to boundary data $H_0$ and $H_1$ satisfying
\bea
\label{BC_delta}
\sigma\,H_0\,\sigma=H_0\,,~~\sigma\,H_1\,\sigma+H_1=\eta\,H_0\,.
\eea
Of course, the case $\eta=0$ is known to correspond to the purely transmitting $\delta$ impurity \ie the system on the full line. We have discussed this case in detail in the 
previous subsection.
To our knowledge, the first analytical study of this problem was performed in \cite{CMR} for the defocusing NLS using a Rosales type expansion of the solution 
\cite{Ros}. The latter can be seen to arise as the Neumann series solution of the Gelfand-Levitan-Marchenko equations appearing in the inverse part of the 
usual ISM method for NLS. The key idea was to formulate appropriate conditions on the initial condition $Q_0(x)$ directly in Fourier space by imposing certain relations on the 
scattering data appearing in the Rosales expansion. In turn, these relations were inspired by the situation in the quantum case where the 
Reflection-Transmission algebras \cite{MRS}
play a role. 

Then, in \cite{HMZ}, important results were obtained for the focusing NLS with a repulsive ($\eta>0$) delta impurity. The initial condition corresponds to a single 
soliton localised on one half-line and the main result concerns the long-time asymptotic behaviour of the solution. The study uses a clever and intricate combination 
of functional analysis estimates method combined with methods of integrable systems like the nonlinear steepest method \cite{DZ}. It is shown that, for high enough velocity, 
the soliton splits into a reflected and transmitted soliton plus radiation. It is our plan to investigate the same problem (and similar ones studied by Holmer and collaborators 
later on) using the method presented here and to compare the results of the two approaches in a future paper.

Finally, in \cite{DP}, the focusing NLS with $\delta$ impurity at small coupling is studied for a special initial condition which has the property of being an even 
function. This allows to map the problem to a scalar problem on the half-line with integrable Robin boundary conditions and use the full power of integrable techniques.
In our setting, this would mean that we choose $Q_0(x)$ such that 
\bea
\label{perm_sym}
\sigma\,Q_0(x)\,\sigma=Q_0(x)\,.
\eea We will show below that this a special case of the notion 
of linearizable initial-boundary conditions that we introduce in the general $N\ge 2$ case.

\subsubsection{NLS with a "jump" defect}	

In the quest for defect/impurity boundary conditions that would preserve the integrability of the model, a privileged class was obtained in \cite{BCZ}. The 
original approach was based on a lagrangian formalism but a key observation was that the obtained defect conditions were frozen B\"acklund transformations at the 
defect location. Using this, in \cite{C}, the author obtained general results on defect conditions and associated generating functionals for the conserved quantities 
for all integrable PDEs in the AKNS scheme \cite{AKNS}. In particular, for NLS, the defect conditions read, in our present notations,
\bea
\begin{cases}
g_1^2(t)+g_1^1(t)=-i\alpha(g_0^2(t)-g_0^1(t))-(g_0^2(t)+g_0^1(t))\Omega(t)\,,\\
\partial_t(g_0^2(t)-g_0^1(t))=\alpha(g_1^2(t)+g_1^1(t))-i(g_1^2(t)+g_1^1(t))\Omega(t)+i(g_0^2(t)-g_0^1(t))(|g_0^1(t)|^2+|g_0^2(t)|^2)
\end{cases}
\eea
where $\Omega(t)=\sqrt{\beta^2+2g|g_0^2(t)-g_0^1(t)|^2}$ and $\alpha,\beta\in\RR$ are two defect parameters.
These defect conditions look very complicated and highly nonlinear. But their origin as B\"acklund transformations of NLS ensure that specific solitonic solutions
can be constructed explicitly. This was done in \cite{CZ} by direct ansatz on the one and two soliton solutions. Using these solutions with $t=0$ and 
$x=0$ as input for the initial and 
boundary data, our approach provides the scheme to compute exactly the scattering data and implementing the inverse part of the method explicitly. 
The final result will of course reproduce the original solutions for $t>0$.

\subsection{Case $N=3$}

In \cite{ACFN}, an important generalization of the study in \cite{HMZ} to the case of focusing NLS on three half-infinite edges connected to a single vertex by 
specific boundary conditions was performed using the same tools and with essentially the same conclusions concerning the splitting of an initial soliton profile 
localised on one of the edges. The boundary conditions used there are part of a general family of boundary conditions that were classified in the context 
of quantum graphs, for instance in \cite{KS}. We show in the next section how these boundary conditions are implemented in our setup for general $N\ge 2$. 
Hence, the particular problem studied in \cite{ACFN} fits in the 
present approach. As already mentioned, a quantitative comparison of their results with results that can be obtained solely by using our approach is 
an important task that we will return to in the future.

\section{$N\ge 2$ Case with General Robin Boundary Conditions}\label{general_case}

In \cite{KS}, a classification of boundary conditions giving to self-adjoint extensions of the Laplacian on a metric graph was derived. 
In the case of the star-graph, these point potentials are well known to induce reflection and transmission between the various edges of the graph. In our 
context, it is natural to try to implement these point potentials as models of local scatterers for nonlinear waves on a star-graph.
The starting point is the linear limit of NLS ($g=0$) which corresponds precisely to the setting of the Laplace operator on a star-graph. Collecting 
the functions $q^\alpha(x,t)$ $\alpha=1,\dots,N$ in a column vector $R(x,t)$, the family of boundary conditions obtained in \cite{KS} is parametrized by $U(N)$, the 
group of $N\times N$ unitary matrices, as follows
\bea
\label{general_Robin}
(U-\1_N)R(0,t)+i(U+\1_N)\partial_x R(0,t)=0\,,~~U\in U(N)\,.
\eea
In the case $N=1$, with $U=e^{2i\alpha}$ this is just $\sin\alpha \,r(0,t)+\cos\alpha\,\partial_x r(0,t)=0$ \ie the Robin boundary condition together with its two limits, 
the Dirichlet and Neumann boundary conditions. We will then call the conditions \eqref{general_Robin} general Robin boundary conditions.

To transfer this to the nonlinear case, we need to rewrite this condition equivalently in the case where the functions $q^\alpha$, $\alpha=1,\dots,N$ are 
collected in a diagonal matrix $Q(x,t)$. Of course, in the linear limit, using $R(x,t)$ or $Q(x,t)$ is equivalent but in the nonlinear case, this changes 
dramatically the form of the interaction term in NLS, and hence the nature of the system, as pointed out in the introduction. To achieve this, we need the following 
simple lemma.

\begin{lemma}
Let $M\in {\cal M}_N$ and let 
\bea
K=\left(\begin{array}{ccccc}
0 & 1 & 0 & \dots & 0\\
0 & 0 & 1 &       & 0\\
\vdots & & \ddots & \ddots & \vdots\\
0 & & & 0 & 1\\
1 & 0 & 0 &\dots &  0
\end{array}\right)\,,~~K^N=\1_N\,.
\eea
Then, there exists a unique decomposition of $M$ on powers of $K$ as
\bea
M=\sum_{j=0}^{N-1}M_j\,K^j\,,
\eea
where $M_j$ is a diagonal matrix for each $j=0,\dots,N-1$.
\end{lemma}
\prf
It suffices to note that 
\bea
M_j=\text{diag}(M_{1,j+1},M_{2,j+2},\dots,M_{N-j,N},M_{N-j+1,1},\dots,M_{N,j})\,.
\eea
\finprf
Let us denote by ${\cal D}_N$ the space of $N\times N$ diagonal matrices over $\CC$. There is a natural isomorphism between $\CC^N$ and 
${\cal D}_N$. Thanks to the previous lemma, we lift this isomorphism to an isomorphism between $End(\CC^N)\cong {\cal M}_N$ and $End({\cal D}_N)$ by defining
\bea
\begin{array}{cccc}
{\cal I}: & {\cal M}_N & \to &  End({\cal D}_N)\\
          &  M    &  \mapsto & \widehat{M}
\end{array}
\eea
where, for all $N\in{\cal D}_N$
\bea
\widehat{M} N=\sum_{j=0}^{N-1}M_j\,K^jNK^{-j}\,,
\eea
and the $M_j$'s are the diagonal matrices appearing in the decomposition of $M$ in powers of $K$. We can now 
define the NLS equation on a star graph with general Robin boundary conditions as the following problem
\bea
&& i\partial_t Q+\partial_x^2 Q-2g\overline{Q}Q^2=0~~,~~
0<x<\infty\,,~~0<t<T\,,\\
\label{BC_star_graph}
&& Q(x,0)=Q_0(x)\,,\\
&&(\widehat{U-\1_N})Q(0,t)+i(\widehat{U+\1_N})\partial_x Q(0,t)=0\,.
\eea
This is a particular case of our general setup where we require $H_0$ and $H_1$ to satisfy the constraint $(\widehat{U-\1_N})H_0(t)+i(\widehat{U+\1_N})H_1(t)=0$. 

\paragraph{Linearizable initial-boundary conditions.}
We want to use the problem with general Robin conditions to introduce and illustrate the notion of linearizable initial-boundary conditions. In Fokas method, there exist 
the so-called linearizable boundary conditions which take their name from the fact that they allow for a solution of the global relation 
by algebraic means only, hence rendering the unified method for IBV problems just as powerful as the ISM for IV problems in linearizing the problem in Fourier space. 
In practice, linearizable boundary conditions correspond to integrable boundary conditions that could be found 
by other methods before the advent of the unified method, like the B\"acklund transformation method initiated by Habibullin \cite{H}. For nonlinearizable boundary conditions, solving the global relation is much more involved and remains essentially a 
nonlinear problem. 

The way to identify linearizable boundary conditions is to exploit natural symmetries of the global relation. The latter represents in Fourier space 
a strong relation between
the initial-boundary data and the integrable bulk dynamics. So far, symmetries of the global relation have been used in such a way as to 
identify boundary conditions which would render the 
problem amenable to solutions for arbitrary initial conditions of the same type as for the problem on the full line. Performing the same reasoning on a star-graph 
essentially leads to trivial linearizable boundary conditions corresponding for instance to disconnected half-lines with Robin boundary conditions. 
However, the example of \cite{DP}
shows that such a restriction on the boundary conditions can be relaxed by restricting one's attention to initial data with a specific symmetry (an even function in 
that case). Indeed, in general the boundary conditions of the $\delta$ impurity are not linearizable but when combined with an even initial condition, one can map the 
problem to a linearizable one.

A posteriori, this is very natural from the point of view of the global relation: one can trade off freedom on the initial data to gain more flexibility 
on boundary conditions leading to IBV problems that can be solved as efficiently as IV problems. In the case of the $\delta$ impurity, the boundary conditions 
\eqref{BC_delta} are invariant by the action of $\sigma$ and so is the bulk dynamics. Therefore, it is natural to split the set of initial data into the two 
eigenspaces of $\sigma$ \ie to 
consider initial data satisfying 
\bea
\sigma\,Q_0(x)\,\sigma=\pm Q_0(x)\,.
\eea
By choosing the initial data in the $+$ subspace, \cite{DP} were able to reduce the problem to a scalar linearizable one, with Robin boundary condition. Note that choosing to use the $-$
subspace would also lead to a scalar linearizable problem but with Dirichlet boundary condition.

We now illustrate this idea for the general Robin conditions above. In this general class, there are distinct representatives that have the additional 
symmetry property that they are invariant by the action of $K$. This is the case for instance for the generalisation of the $\delta$ impurity conditions which read
\bea
\label{general_delta}
K\,H_0\,K^{-1}=H_0\,,~~\sum_{j=0}^{N-1}K^j\,H_1\,K^{-j}=\eta H_0\,.
\eea
More generally, the class of boundary conditions within the general Robin boundary conditions that are such that $U$ appearing in \eqref{BC_star_graph} 
is a circulant matrix, \ie commutes with $K$, allow us to use this extra symmetry.
In that case, it is natural to split the initial data space into the eigenspaces of $K$. The latter is known to have $N$ distinct eigenvalues $\omega^j$, $j=0,\dots,N-1$
where $\omega=e^{\frac{2i\pi}{N}}$ is the $N$-th root of unity.
Therefore, if the initial data satisfies the following symmetry condition
\bea
\label{perm_sym_N}
K\,Q_0\,K^{-1}=\omega^p\,Q_0\,,
\eea
for some $p\in\{0,\dots,N-1\}$,
then the problem can be mapped to a scalar linearizable problem. Note that \eqref{perm_sym_N} is a natural generalization of \eqref{perm_sym}. We show how 
this works for the generalized $\delta$ boundary conditions. They are obtained by choosing $U=\frac{2}{N+i\alpha}J-\1_N$ where $J$ is the 
$N\times N$ matrix with $1$ in every entry and $\alpha$ is a real number related to $\eta$ and representing the coupling. 
The matrix $U$ is circulant and decomposes as
\bea
U=\sum_{j=0}^{N-1}u_j\,K^j\,,~~u_0=\frac{2}{N+i\alpha}-2~~,~~u_j=\frac{2}{N+i\alpha}~~,~~j\neq 0\,.
\eea
Therefore, for the problem with symmetry \eqref{perm_sym_N}, one obtains the following scalar linearizable problem with Robin boundary condition
\bea
(2i\alpha-2(N-1)+\beta_p)\,q(0,t)+i(2+\beta_p)\,\partial_x q(0,t)=0\,,
\eea
where $\beta_p=2(N-1)$ if $p=0$ and $-2$ otherwise, in which case one actually obtains the Dirichlet boundary condition. Here, the function $q$ represents any one of the 
entries of $Q$. This generalizes the 
setup of \cite{DP} and therefore all the methods used there apply here directly. In practice, it means that one can simply study the problem on one of the half-lines of 
the star-graph and the full solution on the complete graph can be reconstructed by applying the symmetry.

Finally, it appears that the special set of initial-boundary conditions considered in \cite{CH}\footnote{We thank D. Noja for bringing this reference to our attention}
for NLS on a so-called $Y$ junction (corresponding to $N=3$ here) corresponds to a weaker version of our notion of linearizable initial-boundary conditions. The authors
considered boundary conditions of the form \eqref{general_delta} for $N=3$ and $\eta=0$ \ie, in our notations,
\bea
K\,H_0\,K^{-1}=H_0~~,~~H_1+K\,H_1\,K^{-1}+K^2\,H_1\,K^{-2}=0\,,
\eea
but with a smaller symmetry constraint on the initial condition. Namely the authors identify two of the edges (called daughter edges in \cite{CH}).  
This reads, in our notations,
\be
\left(\begin{array}{cc}
1 & \\
 & \sigma
\end{array}\right)Q_0(x)\left(\begin{array}{cc}
1 & \\
 & \sigma
\end{array}\right)=Q_0(x)\,.
\ee
As a result, the problem is reduced only to an $N=2$ problem, instead of an $N=1$ problem as in our previous derivation. This problem is then simple enough to 
allow for a study of the Dirichlet to Neumann map in this setting.

\section*{Conclusions}

We introduced a general method that solves the problem of formulating an inverse scattering method for integrable nonlinear equations on a star-graph. The 
key is to map the problem to a matrix IBV problem that can be analysed by a suitable matrix generalization of the unified transform developed by Fokas for 
scalar IBV problems. Although the method was presented for NLS, it is clear that it allows us to tackle any integrable nonlinear equations that can be analysed 
by the Fokas method, that is, any nonlinear equation for which a Lax pair is known.

Our results provide a unifying framework in which one can analyse in great detail the long-time asymptotic behaviour of solutions on a star-graph. This is 
due to the fact that the Riemann-Hilbert approach and the associated nonlinear steepest descent method that are so powerful in the Fokas method, naturally 
extend to our framework. One of our next projects is to implement this program in detail 
in order to compare with the results obtained in \cite{HMZ} and \cite{ACFN} and to go beyond them, hopefully.

In the discussion of the general Robin conditions on the star-graph with $N$ edges, we identified the new concept of \textit{linearizable initial-boundary conditions}
which appear as the natural generalization of Fokas linearizable boundary conditions in the scalar case. The latter turn out to coincide with what was called before "integrable 
boundary" conditions in all known cases. Therefore, the notion of linearizable initial-boundary conditions can be taken as the generalization of the notion of 
integrable boundary conditions in the context of an integrable PDE on a star-graph. The key feature is that one applies the constraints arising from integrability 
not only on the boundary conditions but also on the space of initial conditions. The net result is that certain boundary conditions which would have been declared 
"non integrable" in the usual approach, such as the $\delta$ impurity on the line, can in fact be studied via ISM just as effectively as the traditional 
integrable boundary conditions, provided one works with smaller functional spaces.

In the longer term, the present results open the way to a theory of inverse scattering on arbitrary finite connected graphs. This is because any such graph can be
viewed as a collection of star-graphs connected together by finite edges. Therefore, to complete this program, one will have to combine the present 
approach with the unified transform method applied to finite intervals (see e.g. \cite{BFS}). 

\section*{Appendix} 
\appendix
\section{Proof of Proposition \ref{solution_H1}}\label{AppA}
Let us start by deriving an ``all time version" of the global relation obtained as follows. Let $t\in(0,T]$ and define 
\be
A(t,k)=\overline{(\mu_2(0,t,\bar{k}))_{22}}\,,~~B(t,k)=-(\mu_2(0,t,k))_{12}e^{4ik^2t}\,,
\ee
such that, in particular, $A(T,k)$ and $B(T,k)$ coincide with $A(k)$ and $B(k)$ as defined in \eqref{def_ab}. Inserting these definitions in the block $(12)$ of relation \eqref{def1}
evaluated at $x=0$, we obtain
\be
\label{all_time}
A(t,k)b(k)-B(t,k)a(k)=c(t,k)e^{4ik^2t}\,,
\ee
where $c(t,k)=(\mu_3(0,t,k))_{12}$ can be expressed as
\be
c(t,k)=\int_0^\infty e^{2ikx}Q(x,t)(\mu_3(x,t,k))_{22}\,dx\,,
\ee
and is analytic in $D_1$ and of order $1/k$ as $k\to\infty$. Let us denote $(\mu_2(0,t,k))_{12}$ by $\Phi_1(t,k)$ and $(\mu_2(0,t,k))_{22}$ by $\Phi_2(t,k)$.
Multiply \eqref{all_time} by $ke^{-4ik^2t}a^{-1}(k)$ to get
\be
\label{all_time2}
k\Phi_1(t,k)+ke^{-4ik^2t}b(k)a^{-1}(k)\overline{\Phi_2(t,\bar{k})}=kc(t,k)a^{-1}(k).
\ee
From \eqref{def_wavefunctions2}, we can write
\bea
\Phi_1(t,k)&=&\int_0^te^{4ik^2(\tau-t)}\left[(2kH_0(\tau)+iH_1(\tau))\Phi_2(\tau,k)-igH_0(\tau)\overline{H_0}(\tau)\Phi_1(\tau,k) \right]d\tau\,,\\
\Phi_2(t,k)&=&\1_N+\int_0^t\left[(2k\overline{H_0}(\tau)-i\overline{H_1}(\tau))\Phi_1(\tau,k)+igH_0(\tau)\overline{H_0}(\tau)\Phi_2(\tau,k) \right]d\tau\,.
\eea
Using integration by parts in the first equation, we obtain the following asymptotic representations, as $k\to\infty$ with $k\in D_2\cup D_4$,
\bea
\label{series1}
\Phi_1(t,k)&=&\frac{\Phi^1_1(t)}{k}+\frac{\Phi^2_1(t)}{k^2}+O\left(\frac{1}{k^3}\right)+\frac{e^{-4ik^2t}}{k}C_1+O\left(\frac{e^{-4ik^2t}}{k^2}\right)\,,\\
\label{series2}
\Phi_2(t,k)&=&\1_N+\frac{\Phi^1_2(t)}{k}+O\left(\frac{1}{k^2}\right)\,,
\eea
with $C_1=\frac{i}{2}H_0(0)$ and
\bea
&&\Phi^1_1(t)=-\frac{i}{2}H_0(t)\,,~~\Phi^1_2(t)=-\frac{g}{2}\int_0^t(\overline{H_1}H_0-\overline{H_0}H_1)(\tau)\,d\tau\,,\\
\label{eq_for_H1}
&&\Phi^2_1(t)=\frac{1}{4}H_1(t)-\frac{i}{2}H_0(t)\Phi^1_2(t)\,.
\eea
Applying the residue theorem as well as Jordan's lemma to \eqref{series1} and \eqref{series2}, we obtain
\bea
&&\frac{i\pi}{2}\Phi^2_1(t)=\int_{\partial D_2}(k\Phi_1(t,k)+\frac{i}{2}H_0(t))dk=\int_{\partial D_4}(k\Phi_1(t,k)+\frac{i}{2}H_0(t))dk\,,\\
&&\frac{i\pi}{2}\Phi^1_2(t) =\int_{\partial D_2}(\Phi_2(t,k)-\1_N)dk=\int_{\partial D_4}(\Phi_2(t,k)-\1_N)dk\,,
\eea
where $\partial D_j$ denotes the boundary of the quadrant $D_j$. For $\partial D_2$, the orientation is from $i\infty$ to $-\infty$ via $0$ while 
for $\partial D_4$, it goes from $-i\infty$ to $\infty$ via $0$. This yields, upon noting that integrating over $\partial D_2\cup\partial D_4$ is the same as integrating over $\partial D_1\cup\partial D_3$
\bea
&&i\pi\Phi^2_1(t)=\int_{\partial D_1\cup\partial D_3}(k\Phi_1(t,k)+\frac{i}{2}H_0(t))dk\,,\\
&&i\pi\Phi^1_2(t) =\int_{\partial D_1\cup\partial D_3}(\Phi_2(t,k)-\1_N)dk\,.
\eea
Changing $k$ to $-k$ in the $\partial D_3$ integral in the second equation, this gives
\be
\label{eq_for_p21}
i\pi\Phi^1_2(t) =\int_{\partial D_1}(\Phi_2(t,k)-\Phi_2(t,-k))dk\,.
\ee
Next, we rewrite the first equation as
\be
\label{eq_for_p12}
i\pi\Phi^2_1(t)=2\int_{\partial D_1}(k\Phi_1(t,k)+\frac{i}{2}H_0(t))dk-\int_{\partial D_1}\left[k(\Phi_1(t,k)-\Phi_1(t,-k))+iH_0(t)\right]dk\,.
\ee
The last step is to use the form \eqref{all_time2} of the all time global relation to get
\be
\label{eq_int}
\int_{\partial D_1}(k\Phi_1(t,k)+\frac{i}{2}H_0(t))dk=\int_{\partial D_1}(kc(t,k)a^{-1}(k)+
\frac{i}{2}H_0(t))dk-\int_{\partial D_1}ke^{-4ik^2t}b(k)a^{-1}(k)\overline{\Phi_2(t,\bar{k})}dk\,.
\ee
Finally, note that the analytic properties of of $c(t,k)a^{-1}(k)$ in $D_1\cup D_2$ show that the special combination of functions in the left-hand side 
of \eqref{all_time2} is also analytic and bounded in $D_1$. So that the expansions obtained above for $\Phi_1$ and $\Phi_2$ can be combined and continued
to provide an expansion of $kc(t,k)a^{-1}(k)$ as $k\to\infty$, $k\in D_1$, of the form
\be
\label{expansion_c}
kc(t,k)a^{-1}(k)=\Phi^1_1(t)+\frac{\Phi^2_1(t)}{k}+O\left(\frac{1}{k^2}\right)\,.
\ee 
In fact, the potential obstruction to such a continuation comes from the terms involving $e^{-4ik^2t}$ in \eqref{series1}. 
The previous argument shows that they must vanish in the special combination in the LHS of \eqref{all_time2}. 
It is easy to obtain an explicit check of this fact at order $\frac{e^{-4ik^2t}}{k}$. Using the expansions
\bea
&&a(k)=\1_N+\frac{a^1}{k}+O\left(\frac{1}{k^2}\right)\,,\\
&&b(k)=\frac{b^1}{k}+O\left(\frac{1}{k^2}\right)\,,
\eea
where
\bea
a^1=\frac{ig}{2}\int_0^\infty\overline{Q_0}(x)Q_0(x)dx\,,~~b^1=-\frac{i}{2}Q_0(0)\,,
\eea
we find that the coefficient of $\frac{e^{-4ik^2t}}{k}$ is $C_1+b^1$ which is zero since we assumed $H_0(0)=Q_0(0)$.
Now, \eqref{expansion_c} implies
\be
\label{last_eq}
\int_{\partial D_1}(kc(t,k)a^{-1}(k)+
\frac{i}{2}H_0(t))dk=-\frac{i\pi}{2}\Phi_1^2(t)\,.
\ee 
Combining Eqs \eqref{eq_for_p21}, \eqref{eq_for_p12}, \eqref{eq_int} and \eqref{last_eq} into \eqref{eq_for_H1}, we obtain formula \eqref{form_H1} 
for $H_1(t)$.

\paragraph{Acknowledgments.} We wish to thank N. Cramp\'e for his usefuls comments on the draft of this paper. We are also grateful to the referees for their 
 helfpul and careful reviews.

\end{document}